# Title

# Examining Technology Perspectives of Older Adults with Mild Cognitive Impairment: A Scoping Review

## Author details

Snezna Bizilj Schmidt is a PhD candidate in the School of Information Technology & Systems at the Faculty of Science & Technology, University of Canberra, Bruce, Australian Capital Territory, Australia.
snezna.schmidt@canberra.edu.au (corresponding author)
ORCID ID: https://orcid.org/0009-0004-2415-1206

Stephen Isbel, HScD, MOT, MHA, BAppSc (OT), GCTE, is Professor, Occupational Therapy, School of Exercise and Rehabilitation Sciences, Faculty of Health, and Theme Lead (Innovative Care Models), Center for Ageing Research and Translation, Faculty of Health, University of Canberra, Bruce, Australian Capital Territory, Australia.
ORCID ID: https://orcid.org/0000-0001-5355-3205

Blooma John, PhD is an Associate Professor and Capability Lead in the School of Information Technology & Systems, Faculty of Science & Technology, University of Canberra, Bruce, Australian Capital Territory, Australia.
ORCID ID: https://orcid.org/0000-0001-6725-6025

Ram Subramanian, PhD, is an Associate Professor in the School of Information Technology & Systems, Faculty of Science & Technology, University of Canberra, Bruce, Australian Capital Territory, Australia.
ORCID ID: https://orcid.org/0000-0001-9441-7074

Nathan M. D'Cunha, PhD (Health), BHumNutr (Hons), is Associate Professor, Human Nutrition, School of Exercise and Rehabilitation Sciences, Faculty of Health, and Theme Lead (Dementia and Cognition), Center for Ageing Research and Translation, Faculty of Health, University of Canberra, Bruce, Australian Capital Territory, Australia.
ORCID ID: https://orcid.org/0000-0002-4616-9931

# Abstract

## Background
Mild cognitive impairment (MCI) may affect up to 20% of people over 65. The global incidence of MCI is increasing, and technology is being explored for early intervention. Theories of technology adoption (TA) predict that useful and easy-to-use solutions will have higher rates of adoption, however these models do not specifically consider older people with cognitive impairments, or the unique human-computer interaction (HCI) challenges posed by MCI. There are gaps in understanding the combined impacts of aging and cognitive impairment on factors affecting TA for older people with MCI. It is not clear how MCI impacts HCI and device and interaction modality preferences for this population.

## Objective
This review collates opinions from older people with MCI about technology solutions proposed for them, to understand if solutions are useful, easy to use and what changes are suggested. It also identifies which devices and interaction modalities are preferred, and other factors which may affect usage and adoption.

## Methods
This scoping review was completed according to the Preferred Reporting Items for Systematic Reviews and Meta-Analyses (PRISMA) guidelines. A consistent search was performed across nine electronic databases (ACM Digital Library, EBSCOhost CINAHL Plus with Full Text, EBSCOhost Computers and Applied Sciences Complete, Google Scholar, JMIR, IEEE Xplore, EBSCOhost Medline, Scopus, Web of Science Core Collection) for studies published between 1 January 2014 and 1 May 2024. Extracted data was analyzed using inductive, thematic analysis.

## Results
We identified 4271 studies and after removal of duplicates and screening, 83 studies remained for data extraction. Inductive, thematic analysis of feedback from older people with MCI about technology solutions proposed for them identified five themes (i) purpose and need, (ii) solution design and ease of use, (iii) self-impression, (iv) lifestyle, and (v) interaction modality. Solutions are perceived as useful, even though gaps in functional support exist, however, they are not perceived as entirely easy to use, due to issues related to usability and user experience. Devices which are light, portable, common and have large screens are preferred, as is multimodal interaction, in particular speech, visual/text and touch.

## Conclusions
Using technology can create feelings which positively or negatively affect a user's comfort, confidence and overall well-being. Older people with MCI value independence and autonomy, and solution designs should support these. Usefulness, ease of use, security, privacy, cost, physical comfort and convenience are important considerations for use of technology. Reliable technology creates trust, confidence and feelings of empowerment. This review recommends future work to (i) improve usability and user experience, (ii) enhance personalization, (iii) better understand interaction preferences and effectiveness, (iv) enable options for multimodal interaction, and (v) more seamlessly integrate solutions into users' lifestyles.

# Introduction

Mild cognitive impairment (MCI) affects a person's memory or how they think, feel or behave. It is not a normal part of aging but up to 20% of people over 65 may be affected, and up to 15% of these may progress to dementia [1], [2]. Historically, MCI was deemed as a decline in a person's memory, but the criteria and definition of MCI has expanded with more research to include other cognitive domain deficits and affective attributes [3], [4], [5]. People with MCI may forget things more frequently, lose items more often, struggle to remember words or have problems with language, have trouble making decisions or following instructions, lose their train of thought, have new difficulties regulating emotions, behave more impulsively, have trouble with visual perception, experience physical changes, or be less able to follow daily routines [2]. Given that visual, perceptual, cognitive and behavioral function is needed for human-computer interaction (HCI), the limitations associated with MCI create challenges for HCI, but these can be managed by considered design of technology solutions.

The global prevalence of MCI is increasing and adding stress to both health and aged care services [6]. The number of older Australians with MCI is estimated to grow from 884,000 in 2022 to 2.03 million in 2070 [7], [8]. The Australian Royal Commission into Aged Care Quality and [9] noted aged care services are already under stress and recommended exploring technology options to address the growing needs of older Australians. Australia's National Dementia Action Plan 2024-2034 [10] recommended action be taken to (i) implement evidence-based, early interventions to reduce cognitive decline, and (ii) provide support and resources for people with MCI.

Research has shown that technology provides opportunities to increase healthcare service productivity [11] but only if solutions are adopted by the target users. Aging may result in slower cognitive speed, poorer memory, reduced concentration, and negative impact vision, hearing, speech, dexterity, mobility and learning abilities. These limitations have been found to discourage use of technology [12] and studies have shown that use of digital health solutions is lower in older than younger adults [13]. Additionally, as people experience MCI uniquely, older people with MCI will have different and diverse needs for technology support.

## Literature review

Technology adoption (TA) has been studied from several perspectives. Early models such as the Technology Acceptance Model (TAM) [14], TAM2 [15], and TAM3 [16] take a behavioral perspective and consider TA in an organizational context. They describe TA to be dependent on a user's perception of the technology's usefulness, ease-of-use and the determinants of these. The Unified Theory of Acceptance and Use of Technology (UTAUT) [17] considers TA from a combined behavioral, psychological and information management perspective and UTAUT2 [18] extends UTAUT to consider TA outside of business settings, such as at home, and shows that the context of use has an impact on the relevance and contribution of determinants. However, these models were not developed based on data about older people with cognitive impairment or assessment of healthcare technologies. More recent models such as the Senior Technology Acceptance Model (STAM) [19] and the Model for the Adoption of Technology by Older Adults (MATOA) [20] focus on older users and the effects of physical and cognitive aging on TA. They note that ease of use becomes more important as users age, and MATOA identifies self-management (*the extent to which a person can remain independent and maintain control of their life, emotions and role*) as a determinant for intention to use. The Theory of Life-span Control [21] describes the importance of control and states that people will attempt to retain control of how they behave and interact with their external environment for as long as possible because control contributes to a sense of well-being. It is therefore relevant to consider *user empowerment* as a valuable design feature in technology solutions designed for older people, and this is supported by a literature review of older people's intention to use technologies [22]. STAM and MATOA do not specifically consider TA by older people with cognitive impairments or technology designed for healthcare or assistance with activities of daily living (ADL). The Healthcare Technology Acceptance Model (H-TAM) [23] describes acceptance of healthcare technologies by older people with hypertension. It includes many factors which appear in TAM-based TA models and proposes *compatibility* (how well the technology integrates into a person's everyday life) as a determinant of facilitating conditions.

A study of people's intention to use service robots [24] suggests that TA models may need to be revisited to incorporate the influence of technology innovations such as interactive smart technologies and robots which may be adaptive and exhibit human-like features. These advances have enabled new uses and new modes of interaction, and users can now choose their device and interaction mode, which is empowering. However, it is not clear how this impacts TA for older users with cognitive impairments, in particular perceptions of usefulness and ease of use which have been shown to be determinants of TA. Researchers have investigated effectiveness of unimodal and multimodal interaction for older users and found that modality affects use [25], [26], [27], [28], [29], [30], however these studies did not include older people with cognitive impairments. Subtle speech differences can be observed in people with MCI [31] and literature reviews have reported on the use of voice activated technology solutions by older adults [27], [32], [33]. However, these did not always include older people with cognitive impairment as participants.

There are gaps in understanding the combined impacts of aging and cognitive impairment on factors affecting TA for older people with MCI. It is also unclear how MCI impacts HCI and device and interaction mode preferences. This scoping literature review examines published studies about technology solutions developed for older people with MCI to answer the following research questions:

RQ1: What are the opinions of older people with MCI about the technology solutions which have been proposed for them?
  a) Are they useful?
  b) Are they easy to use or are changes/improvements suggested?

RQ2: What feedback do older people with MCI provide about usage:
  a) Preferred devices?
  b) Interaction modality and ways of use?

To our knowledge, this is the first study which collates opinions and feedback from older people with MCI about technology solutions proposed for them.

# Methods

This scoping literature review was completed according to [34], [35] and is reported according to the Preferred Reporting Items for Systematic Reviews and Meta-Analyses (PRISMA) guidelines [36]. It was pre-registered on Open Science Framework Storage (United States) [37].

## Eligibility criteria

An electronic database search was performed to identify studies published in English between 1 Jan 2014 and 1 May 2024. A 10-year period is considered sufficient given the fast evolution of technology and the potential for change in people's opinions.

## Information sources

The search was performed consistently across nine electronic databases (ACM Digital Library, EBSCOhost CINAHL Plus with Full Text, EBSCOhost Computers and Applied Sciences Complete, Google Scholar, JMIR, IEEE Xplore, EBSCOhost Medline, Scopus, Web of Science Core Collection). The date of the last search is 1 May 2024.

## Search strategy

The search strategy was developed with the assistance of a qualified librarian. The following search query was used in all databases:

> ((MCI OR "mild cognitive impairment" OR “mild cognitive disability” OR “mild neurocognitive”)
> AND
> (technolog* OR ICT OR smart OR wearable* OR computer OR PC OR laptop OR tablet OR “touch screen” OR “touch-screen” OR “mobile phone” OR “mobile device” OR “personal device” OR robot OR reality OR VR OR “assistive technolog*” OR “embodied conversational agent” OR ECA OR multimedia OR “multi-media”)
> AND
> (modalit* OR mode OR channel OR interact* OR engage* OR touch OR type OR voice OR gesture OR use OR usage)
> AND
> (preference* OR experience* OR perception* OR attitude* OR feeling OR practices OR “technology acceptance” OR qualitative OR “design science” OR “innovation resistance theory”))

## Selection criteria

Studies which satisfied the search criteria were imported into referencing software program *EndNote* (Clarivate Analytics, Philadelphia, Pennsylvania, USA) and then uploaded to *Covidence* (Veritas Health Innovation, Melbourne, Australia) to remove duplicates and for screening. *Covidence* is an effective software tool to manage systematic reviews [38].

Three reviewers and the primary author screened the titles and abstracts of each study and excluded any which did not meet the following inclusion criteria:

- Participants have MCI and are living independently in the community, not in residential aged care.
- Participants used or evaluated the technology solutions or devices.
- The study assessed one or more of participant experiences, feelings or attitudes regarding the technology, and preferences for use.

Where it was unclear whether a study met the inclusion criteria, or there was a disagreement among reviewers, the study was carried forward into the full-text review.

Four reviewers and the primary author independently reviewed the full text of the studies extracted from the prior stage based on the following inclusion criteria:

- The study reported on research about participants’ use of technology and their experiences, feelings, attitudes, opinions or preferences.
- If multiple participants cohorts were included, the data were identifiable and reported by cohort.
- Participants were assessed for MCI.
- Participants had no co-morbidities which cause neurological issues or affect cognition.

Any disagreements were resolved by the primary author and approximately 10% of decisions were audited by one of the other reviewers to reduce the risk of bias.

## Data analysis process

Extracted data was captured in a structured Microsoft Excel spreadsheet, designed by the first author, and evaluated before use by two of the other authors to ensure it was fit-for-purpose. Data were extracted about (i) study details (such as author, location of study, year published, journal name, study aims, study type, technology being evaluated, participant details, etc.), and (ii) participant feedback. Participant feedback was recorded as any reference to: (i) a participant opinion about the technology (usefulness, ease of use, usage, context of use), (ii) how the solution made them feel, (iii) preferences for device or interaction, (iv) issues or challenges with use, and (v) recommendations for changes. Statements were captured as written relating to the type of technology and solution.

All extracted data were reviewed for familiarization and to start the process of identifying similarities, relationships and trends. Participant feedback was organized and analyzed using inductive, thematic analysis as described by (Thomas & Harden, 2008), to ensure transparency and traceability from the data to the outcomes. This process has been used effectively to complete other thematic analyses of qualitative data which describe people's experiences, opinions, and feelings.

Inductive thematic analysis of participant feedback was guided by the two research questions and completed in four steps:

1. Keyword Allocation**:** Each feedback item was assigned one or more keyword phrases which described its meaning. The first author used existing keywords or created new ones as needed.
2. Consistency Check**:** Review of all feedback items with the same keyword to ensure consistent allocation.
3. Category Grouping**:** Related keyword phrases were grouped into higher-level categories, and named.
4. Theme Creation**:** The categories were reviewed, and five abstract themes were created to provide meaning to the feedback and answer the research questions.

During the above four-step process, there was a continuous feedback cycle with co-authors.

# Results

A total of 4,271 studies were imported for screening, 687 duplicates were identified and removed by *Covidence,* and 14 duplicates were identified and removed manually. A total of 3,570 studies were screened against title and abstract, and 3,320 studies were subsequently excluded. The full text of 250 studies were assessed for eligibility and 167 studies were excluded, leaving 83 studies for data extraction. Figure 1 illustrates the flow of studies through the PRISMA process and Table 1 lists a summary of the studies and extracted data, sorted alphabetically by the author.
Below, we first list four broad categorizations of the examined studies in terms of (i) study details and nature of participants, (ii) purpose of technological solutions, (iii) types of interaction devices and (iv) interaction modalities, and provide a short description of how studies evaluated usability. This is followed by the resulting themes derived from analysis of participant feedback.

## Categorization of examined study details

### Characteristics of included studies and study participants

Over half of the included studies (n=45) were conducted in Europe with almost a quarter (n=10) of these in Italy. Some of the European studies included participants from multiple countries [39], [40], [41], [42], [43], [44], [45]. Table 1 shows that 61 studies included less than 20 participants, 49 studies included less than 15 participants, 25 studies included less than ten participants, and eight studies included less than five participants. Many studies commented on the lack of cultural diversity among their participants.

Studies used a mixed-methods (n=47), qualitative (n=26) or quantitative (n=10) approach. Studies which did not explicitly declare their type, are classified in this paper based on the type of data that was collected. Qualitative studies used semi-structured interviews, focus groups, workshops and observations to collect data. Quantitative studies used surveys or questionnaires with closed questions to gather data. Most studies, regardless of type, collected quantitative data about participant demographics. Quantitative data about usability, acceptability, and user experience (UX) were typically gathered using questionnaires with responses on a Likert scale.

**Table 1:** Summary of studies, sorted alphabetically by author

| Study details | Study type | Type of technology | Purpose of solution | Participant details |
|---|---|---|---|---|
| Afifi, T. et al., 2023, USA | Quantitative | VR headset | Social interaction / messaging / chat / information provision | Female - 18, Male - 3, Location - community center - 9 had MCI |
| Bartels, S. L. et al., 2020, Netherlands | Mixed methods | Mobile phone | Lifestyle/ADL support, social interaction / companionship | Female - 5, Male - 16, Location - home |
| Beentjes, K. M. et al., 2020, Netherlands | Mixed methods | Touch screen laptop or tablet | Lifestyle/ADL support, social interaction / companionship - helps people find apps for self-management and meaningful activities | Female - 8 (3 in experimental group, 5 in control group), Male - 12 (7 in experimental group, 5 in control group), Location - home |
| Bernini, S. et al., 2023, Italy | Mixed methods | Touch screen laptop or tablet | Cognitive behavior training (CBT) | Female - 5, Male - 5, Location - home |
| Bernini, S. et al., 2023, Italy | Quantitative | PC | Cognitive behavior training (CBT) | Female - 17 (telehealth), 13 (in-person), Male - 14 (telehealth), 12 (in-person), Location - clinic (PC) or home (laptop) |
| Bogza, L. et al., 2020, Canada | Mixed methods | Web / online solution | Decision aid | Female - 6, Male - 6, Location - clinic, research center, or participant home |
| Bouzida, A. et al., 2024, USA | Mixed methods | Robot with integrated touch screen | Cognitive behavior training (CBT) | Female - 1, Male - 2, Location - home |
| Chang, C. et al., 2022, Taiwan | Quantitative | Screen and gesture sensitive device | Physical and cognitive training | Female - 8, Male - 7, Location - clinic |
| Chen, K. et al., 2021, Hong Kong | Mixed methods | Touch screen laptop or tablet | Cognitive behavior training (CBT) | Female - Evaluation - 44, Focus group - 4 (same people), Male - Evaluation - 13, Focus group - 0, Location - clinic |
| Christiansen, L. et al., 2020, Sweden | Qualitative | No device | Data collection e.g. UI design feedback, about technology use or acceptance, icon design preferences, changes to cognition, cognitive training/games design, needs for robotic assistants | Female - 6 (2 were 70 - 75, 4 were over 81), Male - 12 (4 were 70 - 75, 6 were 76 - 80, 2 were over 81), Location - home (15), lab (3) |
| Collette, B. et al., 2021, USA | Qualitative | Integrated virtual assistant | Health or medication management e.g. BP, reminders to drink water | Female - 7, Male - 3, Location - clinic |
| Contreras-Somoza, L. M. et al., 2022, Spain | Qualitative | Touch screen laptop or tablet | Lifestyle/ADL support, social interaction / companionship | Female - 11, Male - 2, Location - clinic |
| Contreras-Somoza, L. M. et al., 2020, Spain, Italy, Greece, Netherlands, Slovenia, France, Serbia, Israel | Qualitative | Touch screen laptop or tablet | Lifestyle/ADL support, social interaction / companionship | Female - 21, Male - 14, Survey - 35 participants (15 Spain, 4 Netherlands, 2 Italy, 2 France, 4 Israel, 5 Serbia, 3 Slovenia) |
| Cunnah, K. et al., 2021, United Kingdon | Qualitative | Web / online solution | Social interaction / messaging / chat / information provision | Total - 100, Location - home - 100 dyads, 49 (control) 51 (intervention) - intervention group got 1:1 training, followed by group training, 25% drop-out, 75 dyads completed study (39 control and 36 intervention) |
| Demiris, G. et al., 2016, USA | Mixed methods | Smart conversational assistant with touch screen (e.g. Google Home, Alexa) | Companionship and reminders (virtual pet) | Female - 10, Male - 0, Location - home |

| Study details | Study type | Type of technology | Purpose of solution | Participant details |
| --- | --- | --- | --- | --- |
| Dixon, E. et al., 2022, USA | Qualitative | No device | Data collection e.g. UI design feedback, about technology use or acceptance, icon design preferences, changes to cognition, cognitive training/games design, needs for robotic assistants | Female - 1, Male - 1, Location - home |
| Franco-Martín, M. A. et al., 2020, Spain | Qualitative | PC | Cognitive behavior training (CBT) | Not stated |
| Gasteiger, N. et al., 2022, Korea, New Zealand | Mixed methods | Robot with integrated touchscreen with pen and magnetic interactive blocks (tactile sensors)<br>Used three robots (Bomy 1, Bomy, and Silbot) in the project | Cognitive behavior training (CBT) | Not stated, Location - location most convenient to participants, including the university, workplaces, clinic, home, or via Skype (for experts) |
| Gelonch, O. et al., 2019, Spain | Mixed methods | Wearable camera | Memory aid - digital camera records daily activities | Female - 4, Male - 5, Location - adult day center |
| Givon Schaham, N. G. et al., 2020, Israel | Mixed methods | Touch screen laptop or tablet | Cognitive behavior training (CBT) | Female - 13, Male - 15, Location - clinic and home |
| Guzman-Parra, J. et al., 2020, Spain, Sweden | Quantitative | No device | Data collection e.g. UI design feedback, about technology use or acceptance, icon design preferences, changes to cognition, cognitive training/games design, needs for robotic assistants | Female - 576, Male - 510, Location - used secondary data - 1086 Dyads 299 with dementia, 787 with MCI - data not split for the MCI and dementia groups |
| Haesner, M. et al., 2015, Germany | Qualitative | No device | Data collection e.g. UI design feedback, about technology use or acceptance, icon design preferences, changes to cognition, cognitive training/games design, needs for robotic assistants | Female - 3, Male - 3, Location - clinic |
| Hassandra, M. et al., 2021, Greece | Mixed methods | VR headset and controller / hand motion trackers / wireless mouse | Physical and cognitive training | Female - 19, Male - 8, Location - clinic |
| Heatwole, S. and Kendra, S., 2022, USA | Qualitative | No device | Data collection e.g. UI design feedback, about technology use or acceptance, icon design preferences, changes to cognition, cognitive training/games design, needs for robotic assistants | Female - 6, Male - 4, Location - outing location |
| Hedman, A. et al., 2016, Sweden | Qualitative | No device | Data collection e.g. UI design feedback, about technology use or acceptance, icon design preferences, changes to cognition, cognitive training/games design, needs for robotic assistants | Female - 2, Male - 4, Location - home |

| Study details | Study type | Type of technology | Purpose of solution | Participant details |
|---|---|---|---|---|
| Hedman, A. et al., 2018, Sweden | Mixed methods | No device | Data collection e.g. UI design feedback, about technology use or acceptance, icon design preferences, changes to cognition, cognitive training/games design, needs for robotic assistants | Female - 18, Male - 19, Location - home - only 21 participants left at year 5 |
| Horn, B. et al., 2023, USA | Qualitative | Mobile phone and smartwatch | Memory aid - facial recognition / aid to identify people | Female - 6, Male - 14, Location - home |
| Hu, H. et al., 2018, Taiwan | Mixed methods | No device | Data collection e.g. UI design feedback, about technology use or acceptance, icon design preferences, changes to cognition, cognitive training/games design, needs for robotic assistants | Female - 18, Male - 13, Location - clinic |
| Infarinato, F. et al., 2020, Italy and Austria, Denmark, Netherlands for participants | Mixed methods | Robot and separate touch screen | Physical and cognitive training | Female - 8, Male - 7, Location - home |
| Irazoki, E. et al., 2021, Spain | Qualitative | Touch screen laptop or tablet | Cognitive training / cognition assessment | Female - 11, Male - 2, Location - clinic |
| Korchut, A. et al., 2017, Poland, Spain | Mixed methods | No device | Data collection e.g. UI design feedback, about technology use or acceptance, icon design preferences, changes to cognition, cognitive training/games design, needs for robotic assistants | Female - 36, Male - 21, Survey |
| Kubota, A. et al., 2022, USA | Qualitative | Robot with integrated touch screen | Cognitive behavior training (CBT) | Female - 0, Male - 3, Location - virtual session |
| LaMonica, H. M. et al., 2017, Australia | Quantitative | No device | Data collection e.g. UI design feedback, about technology use or acceptance, icon design preferences, changes to cognition, cognitive training/games design, needs for robotic assistants | Female - 127, Male - 94, Survey - 137 had MCI, 61 had SCI, 23 had dementia |
| Law, M. et al., 2019, New Zealand | Mixed methods | Robot with integrated touch screen | Cognitive behavior training (CBT) | Female - 6, Male - 4, Location - lab |
| Lazarou, I. et al., 2021, Greece | Mixed methods | No device | Data collection e.g. UI design feedback, about technology use or acceptance, icon design preferences, changes to cognition, cognitive training/games design, needs for robotic assistants | Female - 9, Male - 6, Survey |
| Leese, M. et al., 2021, USA | Mixed methods | Fitness tracker smartwatch | Collect daily movement/exercise data | Female - 1, Male - 14, Location - home |
| Lindqvist, E. et al., 2018, Sweden | Qualitative | No device | Data collection e.g. UI design feedback, about technology use or acceptance, icon design preferences, changes to cognition, cognitive training/games design, needs for robotic assistants | Total - 5, Location - clinic |

| Study details | Study type | Type of technology | Purpose of solution | Participant details |
|---|---|---|---|---|
| Madjaroff, G. and Mentis, H., 2017, USA | Qualitative | No device | Data collection e.g. UI design feedback, about technology use or acceptance, icon design preferences, changes to cognition, cognitive training/games design, needs for robotic assistants | Female - 3, Male - 2, Location - clinic |
| Maier, A. M. et al., 2015, Denmark | Qualitative | Wearable watch, smartphone and smartboard | Memory aid for daily activities / cognitive support or rehabilitation / share health information | Total - 6, Location - clinic - 3 in focus group, 3 in usability evaluation of 2 prototypes - prototype 1 used a smartphone (not wearable technology), prototype 2 used a smartphone and a smartboard (wall calendar) |
| Manca, M. et al., 2021, Italy | Mixed methods | Robot with integrated touch screen | Cognitive behavior training (CBT) | Female - 9, Male - 5, Location - lab |
| Manera, V. et al., 2016, France | Quantitative | VR headset and controller / wireless mouse AND screen and gesture sensitive device, 3D glasses | Cognitive behavior training (CBT) - train selective and sustained attention using image based rendered environment | Female - 13, Male - 15, Location - lab |
| Mathur, N. et al., 2022, USA | Mixed methods | Smart conversational assistant with touch screen (e.g. Google Home, Alexa) | Health or medication management e.g. BP, reminders to drink water | Female - 4 total - phase 1 - 2, phase 2 - 2, Male - 8 total - phase 1 - 5, phase 2 - 3, Location - home |
| Matsangidou, M. et al., 2023, Cyprus | Mixed methods | VR headset | Regulate emotions and mood | Female - 19, Male - 11, Location - clinic |
| Mattos, M. K. et al., 2021, USA | Mixed methods | PC and smartwatch | Insomnia intervention / sleep management | Female - 7, Male - 5, Location - home - only 10 completed the study |
| McCarron, H. R. et al., 2019, USA | Mixed methods | Mobile phone and smartwatch | Memory aid - facial recognition / aid to identify people | Female - 25, Male - 23, Location - home - 29 participants had dementia |
| Mehrabian, S. et al., 2014, France | Qualitative | Touch screen laptop or tablet | Lifestyle/ADL support, social interaction / companionship | Female - 19, Male - 11, Location - lab |
| Mondellini, M. et al., 2022, Estonia | Mixed methods | VR headset and controller / hand motion trackers / wireless mouse | Cognitive training / cognition assessment | Female - 14, Male - 1, Location - lab |
| Moro, C. et al., 2019, Canada | Mixed methods | Humanoid robot, cartoon robot, tablet | Lifestyle/ADL support | Female - 6, Male - 0, Location - kitchen |
| Mrakic-Sposta, S. et al., 2018, Italy | Mixed methods | VR and stationary bike with controller on handlebars | Physical and cognitive training | Female - 6 - 3 in experimental group, 3 in control group, Male - 4 - 2 in experimental group, 2 in control group, Location - lab |
| Nie, Q. et al., 2020, USA | Mixed methods | Web / online solution, Webcam | Social interaction / messaging / chat / information provision | Total - 7 - 5 in phase 1 and 2 in phase 2, Location - clinic |
| Nieto-Vieites, A. et al., 2024, Spain | Mixed methods | Touch screen laptop or tablet | Cognitive behavior training (CBT) | Female - Study 2 - 5, Study 3 - unknown, Male - Study 2 - 3, Study 3 - unknown, Location - clinic |
| Ortega Morán, J. et al., 2024, Spain, Portugal | Qualitative | Touch screen laptop or tablet | Cognitive behavior training (CBT) | Female - 16, Male - 3, Location - clinic |
| Park, C. et al., 2022, USA | Quantitative | Touch screen tablet and wearable motion sensor device | Physical and cognitive training | Female - 12, Male - 2, Location - home |
| Park, J. et al., 2020, Korea | Quantitative | VR headset and controller / hand motion trackers / wireless mouse | Cognitive behavior training (CBT) | Female - 7 - Control group, 7 - VR group, Male - 3 - Control group, 4 - VR group, Location - clinic |
| Piasek, J. et al., 2018, Poland | Mixed methods | Robot with integrated touch screen | Physical and cognitive training | Female - 3, Male - 1, Location - home |
| Pino, M. et al., 2015, France | Mixed methods | Robot with integrated touch screen | Lifestyle/ADL support, social interaction / companionship | Female - 6, Male - 4, Location - clinic |

| Study details | Study type | Type of technology | Purpose of solution | Participant details |
|---|---|---|---|---|
| Quintana, M. et al., 2020, Sweden, Spain | Mixed methods | Touch screen laptop or tablet with tablet pen | Memory aid for daily activities / cognitive support or rehabilitation / share health information | Female - Sweden - 3<br>Spain - 5, Male - Sweden - 6, Spain - 5, Location - clinic and home |
| Rossi, S. et al., 2024, Italy | Mixed methods | Robot with integrated touch screen | Lifestyle/ADL support | Female - Phase 1 - 2, phase - 3, Male - Phase 1 - 2, phase - 4, Location - phase 1 - lab, phase 2 - home |
| Saini, J. et al., 2018, USA | Mixed methods | Video conferencing (Zoom), Webcam | Cognitive behavior training (CBT) | Total - 12, Survey - 6 in each randomized group (face-to-face CBT or CBT via video conferencing) |
| Scase, M. et al., 2018, Italy | Qualitative | Touch screen laptop or tablet | Cognitive behavior training (CBT) | Female - Development - 3 focus groups - group 1 - 4F, group 2 - 4F, group 3 - 3F, Evaluation - 22F<br>Male - Development - 3 focus groups - group 1 - 5M, group 2 - 1M, group 3 - 1M, Evaluation - 3M<br>Location - clinic and home |
| Scheibe, M. et al., 2021, Germany | Qualitative | TV with remote control or tablet AND sphygmomanometer (to measure blood pressure) | Telemonitoring medical app to communicate with multi-disciplined health professionals | Female - 8, Male - 4, Location - home |
| Shamir, D. et al., 2024, Israel | Mixed methods | Touch screen laptop or tablet | Cognitive behavior training (CBT) | Female - 6, Male - 8, Location - home |
| Shellington, E. M. et al., 2017, Canada | Mixed methods | Mobile phone | Cognitive behavior training (CBT) | Female - 14, Male - 5, Location - home |
| Shin, M. H. et al., 2022, USA | Qualitative | Remote controlled robot (robot is not with user, but controlled via app on PC, tablet or mobile phone | Lifestyle/ADL support | Female - 0, Male - 6, Location - clinic or home |
| Stogl, D. et al., 2019, Germany | Quantitative | Robotic walker - user holds handles and pushes device to exercise | Physical and cognitive training | Female - 2, Male - 8, Location - clinic |
| Stramba-Badiale, C. et al., 2024, Italy | Mixed methods | VR (or TV), joystick and foot pad/rudder | Assist with navigation and training of spatial memory | Female - 4, Male - 3, Location - lab |
| Tuena, C.et al., 2023, Italy | Mixed methods | VR (or TV), joystick and foot pad/rudder, 3D glasses | Assist with navigation and training of spatial memory | Female - 2, Male - 6, Location - lab |
| Van Assche, M. et al., 2024, Belgium | Qualitative | Robot with integrated touch screen | Data collection and design guidance for robot use cases | Female - 16, Male - 14, Location - home |
| Votis, K. et al., 2015, Greece | Qualitative | Touch screen laptop or tablet | Cognitive behavior training (CBT) | Total - 17, Location - lab |
| Wargnier, P. et al., 2018, France | Mixed methods | Smart conversational assistant with touch screen (e.g. Google Home, Alexa) | Health or medication management e.g. BP, reminders to drink water | Female - 11, Male - 3, Location - clinic - 5 had AD |
| Weering, M. et al., 2019, Netherlands | Qualitative | Web / online solution | Physical and cognitive training | Female - 43, Male - 14, Location - clinic |
| Wolf, D. et al., 2019, Germany | Mixed methods | VR headset | Lifestyle/ADL support | Female - 6, Male - 0, Location - clinic kitchen |
| Wu, Y. et al., 2016, France | Qualitative | No device | Data collection e.g. UI design feedback, about technology use or acceptance, icon design preferences, changes to cognition, cognitive training/games design, needs for robotic assistants | Female - Focus group 4, interview 12, Male - Focus group 1, interview 3, Location - lab |
| Wu, Y. et al., 2014, France | Mixed methods | Robot with integrated touch screen | Lifestyle/ADL support | Total - 6, Location - clinic |

| Study details | Study type | Type of technology | Purpose of solution | Participant details |
|---|---|---|---|---|
| Yamazaki, R. et al., 2021, Japan | Qualitative | Robot - sitting on table, programmed to interact with user, sing, talk and respond to spoken words | Companionship | Female - 2, Male - 0, Location - home |
| Yun, S. J. et al., 2020, Korea | Quantitative | VR headset and controller / hand motion trackers / wireless mouse | Cognitive behavior training (CBT) | Female - 6, Male - 5, Location - clinic |
| Yurkewich, A. et al., 2018, Canada | Mixed methods | Touch screen laptop or tablet | Social interaction / messaging / video chat / information provision | Female - 6, Male - 2, Location - home |
| Zafeiridi, P.et al., 2018, Italy, Spain, France, United Kingdom | Mixed methods | Web / online solution | Lifestyle/ADL support, social interaction / companionship | Female - 14, Male - 10, Location - home |
| Zedda, E. et al., 2023, Italy | Mixed methods | Robot with integrated touch screen | Cognitive behavior training (CBT) | Female - 6, Male - 10, Location - clinic |
| Zhang, B. and Gao, Y., 2023, China | Mixed methods | No device | Data collection e.g. UI design feedback, about technology use or acceptance, icon design preferences, changes to cognition, cognitive training/games design, needs for robotic assistants | Total - 5, Location - clinic |
| Zhang, Q. and Liu, Y., 2022, China | Mixed methods | Touch screen laptop or tablet with foot stand with pressure sensors | Physical and cognitive training | Female - 79, Male - 58, Location - survey and usability evaluation in lab - 137 responses for phase 1 (requirements gathering) but only 30 responses were from people with MCI - 107 responses were from family of people with MCI |
| Zhu, D. et al., 2024, China | Mixed methods | Mobile phone | Memory aid - digital story telling | Female - 9, Male - 3, Location - clinic |
| Zubatiy, T. et al., 2021, USA | Mixed methods | Smart conversational assistant with touch screen (e.g. Google Home, Alexa) | Lifestyle/ADL support | Female - 4, Male - 6, Location - home - MCI/caregiver dyads included in study |

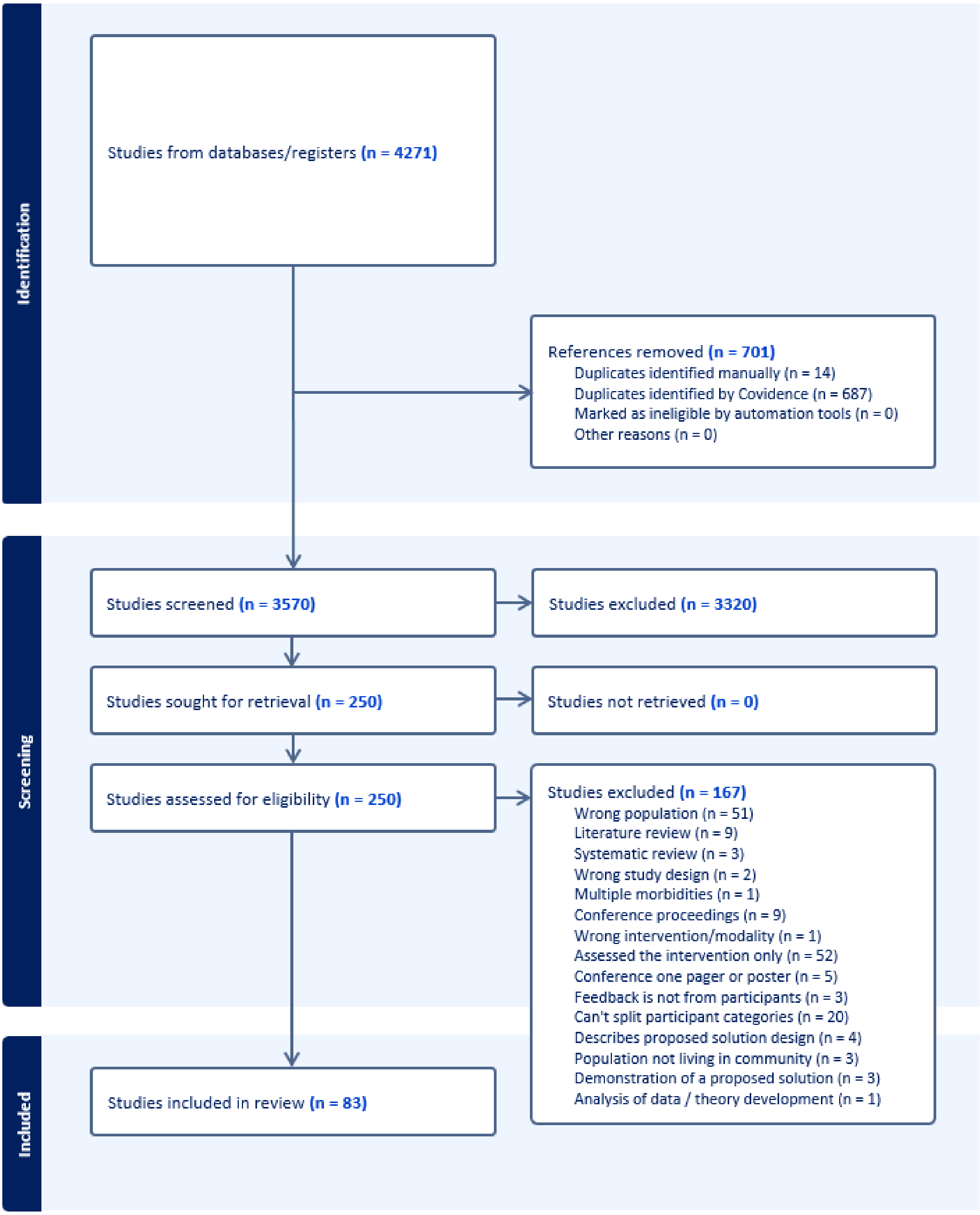

**Figure 1** – Flow of studies through the PRISMA process

## Purpose of technology solutions in studies

Over a third (n=32) of solutions provide cognitive training in the form of cognitive behavioral therapy (CBT) (n=21), a combination of CBT and physical training (n=9) or CBT and cognitive assessment (n=2). Other functions mentioned as needs by participants and supported by solutions were lifestyle/ADL support (n=13), memory training/management (n=6), health and medication management (n=4), staying in touch with family/friends, social engagement, entertainment (n=4), companionship (n=2), spatial navigation/orientation (n=2), insomnia intervention (n=1) and daily exercise monitoring (n=1). Solutions also provided support for decision support (n=1) and mood regulation (n=1), but these were not identified as needs by participants. Further needs identified by participants but not supported by solutions were management of finances and help to locate lost items. Sixteen studies were designed to collect data to (i) guide design of future solutions, (ii) inform amendments for existing solutions, or (iii) collect information about needs of people with MCI.

### Interaction devices

Studies included various technology platforms and devices as illustrated in Table 2 in the Appendix. Almost half (n=37) of the proposed solutions include a touch screen as a component of a laptop or tablet (n=17), a robot (n=14) or a mobile phone (n=6). This reflects research by [25], [46] which showed that touch screen devices are effective and easy to use by older people. Some (n=4) robot solutions did not include sound, but the majority (n=12) included voice or sound. Eleven solutions involved the use of Virtual Reality (VR). Eight solutions included wearables (other than VR headsets or 3D glasses) in the form of a smartwatch (n=5), a wearable motion sensor (n=1), a sphygmomanometer to measure blood pressure (n=1) and a wearable camera (n=1). Six studies included a mobile phone, among which 3 paired the phone with a smartwatch. Five studies used a conversational agent such as Google Home (n=3), an integrated conversational agent on a laptop (n=1) and a virtual pet avatar on a laptop or tablet (n=1).

### Interaction modalities

The variety of technology and devices made different options available for users to interact with the solutions and vice versa. Over half the studies (n=48) involved touch interaction either as unimodal (n=24) or multimodal (n=24). Seven of these touch solutions could also be used with a keyboard and mouse. This reflects research by [25], [46] which demonstrates that touch screen interaction is an effective interaction mode for both old and young users even with some usability issues. Nine studies combined a touch screen device with speech output, while 13 studies combined a touch screen device with speech input and output. Combining visual and speech modalities is supported by research findings that multimodal interaction is more effective and usable for older adults [47], as compared to unimodal touch or speech [48]. Voice or sound was associated with 29 solutions; 13 studies involved solutions which provided speech/sound outputs, and 16 studies involved solutions which afforded speech/sound input and output. Voice or sound outputs were associated with VR solutions (n=4), robots (n=6), tablet device solutions (n=2) and a web/online solution (n=1). Voice or sound input-plus-output was associated with VR solutions (n=2), robots (n=6), tablet device solutions (n=3), embodied conversational agent (ECA) solutions (n=3), web/online solution (n=1) and a mobile phone app (n=1). Only two studies used gesture as an interaction modality; one used a robot which gestured to the user as a demonstration of extraversion [49], and the other was a cognitive game based on Tetris which included a user gesture capture device [50]. Four studies used haptic (vibration) interaction via a smartwatch and a touch screen device (n=2), a vibrating keyboard with a touch screen device (n=1) or a vibrating handset as part of a VR solution (n=1). Additional interaction modalities provided by VR solutions included joysticks and pedals/rudders (n=4) and robot solutions with lights, facial expressions and movement (n=7). Table 2 in the Appendix lists the device and interaction modality combinations for each study.

### How studies evaluated usability

Studies used different methods to evaluate usability. Some studies used the System Usability Scale (SUS) [45], [50], [51], [52], [53], [54], [55], [56], [57], [58], [59], [60] which is a simple, ten-item Likert scale used to assess usability [61]. Research [62] has shown that the SUS is suitable for evaluating digital health application. Some studies of robotic solutions used the Almere questionnaire [63], which is based on UTAUT and developed to evaluate acceptance of assistive social agents by older people [64], [65], [66], [67], [68], [69]. Several studies designed their own questionnaires based on TAM, TAM3, STAM, MATOA or UTAUT.

## Themes derived from thematic analysis of participant feedback

Inductive, thematic analysis of extracted data identified five themes (i) purpose and need, (ii) solution design and ease of use, (iii) self-impression (how does the solution make the user feel), (iv) lifestyle (how well the solution fits with the user's life and routines), and (v) interaction modality.

### Purpose and need

Participants identified that improving or maintaining their existing cognition level is the priority need, and over a third of solutions focused on this function. The remaining studies reveal other needs identified by participants, and only two studies provided solutions for activities which were not identified as needs. Assistance with managing personal finances was identified as a need [70], [71]; however, none of the studies provided a solution to this end. Being able to manage one's own finances decreases vulnerability and is an important enabler for independence. Further investigation is needed to determine how to support older people with MCI to have control of and manage their finances in a safe and secure manner. Participants also identified help to find lost items as a need [69], which was again not supported by any studies. The answer to RQ1a is that proposed solutions for older people with MCI are perceived as useful, but there are also gaps where functional support is missing. Given the demonstrated importance of perceived usefulness for TA, this is an area where further work is needed.

### Solution design and ease of use

Study participants reported a variety of usability and usage issues and noted that an ability to configure and personalize a solution provides flexibility and increases usability.

> *"There's all grades of dementia as well isn't there. And some would need more help than others"* [72].

Participants said it would be useful if they could adjust (i) solution tasks to match user skills [56], [73], [74], [75], (ii) task complexity [51], [58], [72], [76], (iii) task workflow [77], (iv) timing of reminders [78], (v) user interface design (colors of background, text and widgets [72], [79], size of text and widgets [80], [81], color contrast [72]), (vi) speed, volume and accent of speech [45], [72], (vii) language formal/casual [64], (viii) robotic facial features and height [41], [66], [67], [68], and (ix) avatar image [82].

Consistent design, labelling and widget placement were mentioned as important enablers of usability [59], [83] and this is especially critical for users with cognitive impairments or memory concerns. Participants added that they may use multiple toolsets and that seamless integration and consistency across these toolsets would increase usefulness and ease of use.

Participants provided ideas about how to increase usability of solutions, such as (i) reduce or eliminate the need to remember passwords or specific trigger words because this increases complexity [84], (ii) make messages specific and include sufficient detail so the message is actionable [57], (iii) provide sufficient time to read text messages [55], [56], [85], (iv) use language appropriate for older users [79], (v) use simple, familiar and common words [45], [59], (vi) use sound as well as visual indicators to make it obvious that an action was performed successfully [45], (vii) avoid loud noises and too many animations because these create a distraction [75], and (viii) make task steps simple [86]. Participants prefer solutions which provide immediate feedback about task progress and task completion, because they are engaging and provide motivation and incentives for use [44], [57], [75], [87], [88]. This illustrates how UX can influence TA.

Specific feedback about user interfaces were (i) layout should be simple and not too busy or overwhelming so that controls are obvious and the interface does not overstimulate or place a burden on the user's cognition [52], [59], [83], [85], (ii) use visual aids [89], (iii) use simple images so the item is easily recognizable [80], [90], (iv) use intuitive and simple navigation [91], [92], (v) use bright and contrasting colors [79], [91], [93], (vi) use large text, buttons and widgets [59], [91], [94], (vii) ensure all options are visible so information is transparent [52], and (viii) list options in order of user priorities because users often select the first option listed [52]. These comments are consistent with design principles proposed by [95], [96], [97].

Reliable technology creates trust, confidence and feelings of empowerment [45], [68]. Studies where participants experienced technical issues [40], [64], [72], [98] or if the device had a low battery life or charged slowly [99] resulted in frustration and negative feedback.

Participants commented that age-related physical limitations may affect how they use technology and so accessibility is an important design feature of technology solutions for older people with MCI [45], [79], [84], [94]. Context of use and location were also raised as considerations in this regard (e.g. rural vs metropolitan, speed of internet connection [76]).

Studies showed that older people with MCI prioritize security and privacy [68], [100]. Some participants said they prefer robots to have mechanical features and to be shorter than a person [66], [68] because they felt less under surveillance. This preference has also been reported by other research [101]. Safety was identified as a priority, related to both (i) personal safety when using technology (especially robots or VR solutions), and (ii) online activity such as internet and social media use [100]. Participants said that even though privacy is a need, they are willing to compromise privacy and share personal data or their location with medical professionals or carers in an emergency [102]. Some participants said that it would be useful to receive regular reminders of what data is being shared [43].

Given the breadth of usability issues reported, and the number of changes suggested, the answer to RQ1b is that proposed solutions for older people with MCI are not perceived as completely easy to use, and this is an area where future technologies need to focus.

### Self-impression

Participants commented that the feelings they experience while using a solution can positively or negatively affect their comfort, confidence and overall well-being. The wearable camera solution made participants feel uncomfortable because it drew attention to them and they felt conspicuous.

> "*if people don't know you, they look at you as if they are thinking well, what is he doing?*" [103].

Participants also felt vulnerable because the camera may capture private moments, such as when they are going to the bathroom.

> *"the camera is black and people look at it a lot. I think that if it were a more natural color, people would look at it less"* [103].

Some participants said that technology would be useful for other people, but not themselves. Technology is often associated with negative aspects of aging such as loneliness or ill-health [104] because it advertises the user's limitations and lack of independence.

> "*The telecare system would be useful for me if I had more deficits. But so far, I can manage by myself at home*" [105].

Participants said they want solutions to reflect their identity and present a positive self-image. They suggest that solutions should (i) use affirming language which does not remind them about their cognitive limitations [71], [85], [94], [98], (ii) include non-directive task workflows (e.g. participants prefer a system to check-in rather than remind [57]), and (iii) not draw attention to the user or the solution, which is especially true for robotic solutions [66], [68], [104] and the wearable camera [103]. Self-concept and social presence have been shown to be determinants of TA in the MATOA model, and a literature review [106] describes similar feedback.

Participant personality traits did not appear to affect acceptance of robots in the long term [65], however, differences were observed over the study duration. Some participants felt comfortable engaging with a robot and interacted with it in a human-like manner [67], [107]. They said a robot is novel and suggested that it should be human-like with autonomy, be able to perform simple tasks and interact via two-way communication [41], [51]. This is consistent with existing research [108] which demonstrated that robots which take initiative, exhibit more activity, and are extraverted, are better accepted by users.

> "*I don't see much difference between [my doctors] and the robot except one is ran by electricity and the other one is a human being*" [51].

Other participants said they want to engage with real people rather than anthropomorphic technology. They value human connection [59], [104], [107] and there is a concern that use of technology will reduce their existing human contact. These participants preferred more machine-like robots which did not imitate human traits.

> "*A robot doesn't have a heart*", "*It must be for people who are very handicapped. It's not for me. It makes me think that my life is terminated. I'd rather die than live with a robot*" [66].

Other concerns identified by participants relating to the use of robot solutions were that the robot would (i) be hard to use or experience technical problems, (ii) not do what they needed/wanted it to do, (iii) be expensive, (iv) increase isolation, or (v) make them too dependent on the robot [66], [68], [104]. Nevertheless, many participants felt optimistic about anthropomorphic solutions and recognized their potential

> "*The human aspect, the "feeling", is hard to create with a robot. But there is an enormous progress in that world, and I do believe in it. Of course, still in combination with us, humans*" [107].

Participants who said they prefer more machine-like robots were less willing to create an emotional connection with the robot [66], [104]. However, some participants got very attached to their robot solutions, treated the robot like a friend, and became sad when the study ended [41], [109], [110], [111]. This raises important issues for designers about how to support users if technology solutions fail or become unavailable.

## Lifestyle

Participants said that independence and autonomy are important, and some said they felt a responsibility to their relatives and carers to maintain their self-sufficiency [51]. Participants want to feel in control of their lives [42], [51], [57], [59], [70], [71], [75], [104], [112] and technology solutions which create a sense of empowerment are preferred rather than solutions which impose mandatory interventions or training [75]. While participants appreciate the increased independence technology can provide [71], they are also concerned about receiving too much assistance (from technology or people) in case they become reliant on this and lose their autonomy [104] or independence [41], [71]. The studies reinforced that IT literacy, training and availability of support from carers and family are enablers for acceptance and on-going use of technology [81], [88], [100] which is consistent with technology acceptance models. However, it is important that technology solutions do not create a dependence on support services which in turn would erode independence.

Cost was identified by study participants as an important factor when considering adoption of a technology solution [66], [68], [88], [105], [107]. This is consistent with UTAUT2 and H-TAM which include cost, price/value or perceived benefit as determinants of adoption.

Physical comfort and convenience are important considerations for the use of technology. Most participants enjoyed the novelty of VR solutions; however, some experienced visual discomfort or nausea, found the headsets uncomfortable, or said the VR solutions were not practical [58], [90], [113].

> "*Well, I liked it, but it was too heavy and big for me. I had difficulty breathing because it wasn't staying in place and fell into my face blocking my nose*" [58].

The answer to RQ2a is that participants prefer devices which are light and portable (mobile phone or tablet), commonly owned (mobile phone), readily available, and have a reasonably large screen [87], [105], [114], [115]. This review also identified that older people with MCI do not like small wearable technology such as watches, because the screens are too small, and many participants said that watches are not convenient because they are no longer part of their everyday toolset [87]. In addition, the physical size of the technology solution and its placement in the home, affects ease of access and contributes to convenience (e.g. smaller robots take up less room and are less intrusive in a home setting [51], [107]).

Studies where solutions integrated into participants' existing lifestyles and routines made participants feel comfortable, confident and more in control. This is consistent with the STAM, MATOA and H-TAM models, which show that compatibility, ease of use and maintaining a sense of control are important for older users of technology.

> *"I like that Google kind of feels like a part of the house, and not something that I have to keep answering to all the time like my morning alarm"* [116].

## Interaction modality

Study participants provided positive feedback for voice interaction and said it was intuitive and easier than writing [51]. They also mentioned challenges such as (i) the solution's accent was hard to understand or the solution did not understand the user's accent or speech pattern [72], [98], (ii) the speech was too fast or the volume was too loud/soft [85], [109], (iii) the solution spoke for too long [64], (iv) the solution only recognized a limited vocabulary and participants had to remember specific 'trigger' words to activate the solution or give it commands [84], [90], [110].

> "*If it was more New Zealand English spoken it would be easier for me I think*" [72].

Similar feedback has been documented by other researchers studying voice interaction with older users [27], [33]. Research has demonstrated that subtle differences in speech can be observed in older people with MCI [31], and people with cognitive decline speak slower, in smaller chunks, pause more often, have longer silences, respond slower to questions, and provide shorter answers [117]. This highlights the importance of including older people with cognitive impairments when designing and evaluating voice interaction solutions. Mozilla [118] is leading the world's largest open-source voice data project called 'Common Voice'. The project is building and refining a voice dataset which is being used to teach machines to speak. To date, it includes 131 languages and 32,584 hours of recorded voice data, however analysis [119] shows that people over the age of ninety are represented only in the Mongolian and Polish language and people in their eighties are only represented in the Esperanto, Turkish, Abkhazian, Toki Pona and Hebrew languages. Of the English contributing voices, there are 0.01% in their seventies, 0.04% in their sixties, 0.05% in their fifties, 0.09% in their forties, 0.14% in their thirties, 0.25% in their twenties, 0.06% in their teens and 0.36% of unspecified age. It is not known if any of the recorded voices are from older people with MCI. Technology solutions utilizing speech interaction increase their risk of error if they cannot recognize the speech of older users or speech of people who are cognitively impaired. Users' current optimism creates an opportunity to invest in voice interaction to improve its usability and effectiveness.

Participants said that touch interaction is effective and easy, especially for less IT literate people [76], [91], but they experience difficulties such as (i) trouble dragging items across the screen, (ii) knowing how much pressure to use [120], (iii) small screen/widget size making it hard to select the target accurately, and (iv) using the touch screen if the user has dexterity problems [45], [92], [98], [120]. Some participants chose to use an instrument to avoid these challenges [45], [92].

> *"I have a problem only with, first, with the gentleness of the finger. My finger isn't the gentlest thing"* [120].

These findings are consistent with outcomes from a literature review investigating use of touch screen use by older users [46], and an investigation of touch interaction which identified usability issues associated with clicking, dragging, zooming, and rotating on a touch-device for both older and younger users [25]. Based on feedback from study participants in this review, it appears that these challenges have not been fully resolved.

Participants said robotic assistants should be able to engage in both a verbal and non-verbal manner, and they prefer robots which can speak, listen and answer questions rather than those with one way interaction via a touch screen [41], [51]. If the solution is embodied, they want to be able to touch it physically and interact with it rather than simply view it on a screen [110]. Active robots which performed tasks were deemed more pleasant [65]. This is consistent with findings that robots with unimodal auditory two-way feedback are highly usable and acceptable for older people [27].

Participants were agnostic to haptic interaction, except for conveying that a smartwatch was not a preferred device and suggested that it is useful to be able to control technology remotely [64], [86] which supports the need for comfort and convenience.

The answer to RQ2b is that participants prefer multimodal interaction, in particular speech, visual/text and touch [57], [59], [75], [77], [84], [121] because it is flexible, provides options for use if people have physical or cognitive limitations, and makes solutions accessible to people with poor hearing or vision/speech limitations. This supports existing research about effectiveness of different interaction modalities for older users which showed that multimodal interaction is most effective [28], [29], [122] and that effectiveness of interaction modalities and multi-modalities is task and user dependent [26], [29], [123]. However, no studies include older people with cognitive impairment which creates a gap in understanding the effectiveness of different interaction modality combinations for this population.

# Discussion

## Principle findings

Participants across studies reported similar needs but often in a different priority order, and this scoping review is not able to (i) provide a definitive priority order for needs, (ii) identify if needs are relatable to technologies or devices, or (iii) attribute differences to the context of use, cultural differences, participant demographics, or the way MCI is affecting participants' lives. Mondellini et al. [124] found that user needs appear to be location-specific but further research is required to understand which needs are location-specific versus those which are generalized.

Fadzil et al. [125] completed a review of existing technology solutions for older people with cognitive impairments and found that current solutions are focused on (i) caregiver support, (ii) strengthening cognition, (iii) cognitive rehabilitation, and (iv) using technology for disease prevention or self-management. Fadzil et al. [125] does not comment on reported needs, technology preferences, or opinions about existing solutions of older people with cognitive impairments. Neither does it describe how well existing technology supports user needs, or how well technology is aligned to users' preferred ways of doing things. This gap contributes to the lack of understanding of TA by older people with MCI.

Our review identified that security and privacy are important for older people with MCI, nonetheless they are prepared to compromise these to feel safe. However, they do not want to be continuously watched or feel like they are under surveillance. Nastjuk et al. [126] used the term 'techno-invasion' to describe the stress caused by constant observation and monitoring. Solutions must balance the need for users' privacy and security with their need for safety and well-being.

A positive UX is a significant enabler for technology acceptance, and ease of use contributes greatly to UX, which suggests that designers must give both priority. User attitude has been identified in technology acceptance models as a determinant of use, and several studies commented that MCI has an impact on a person's attitude to technology use [87], [100]. This observation has also been made by other researchers [127] who noted that we need greater understanding about how sociodemographic factors and cognition affect attitude.

Some older people with MCI are not willing to connect with a robot because they prefer to interact with a human, while others are comfortable engaging with anthropomorphic solutions. Our review identified that participants who created an emotional connection with an anthropomorphic solution were vulnerable to becoming depressed once the solution was no longer available. This raises important issues for designers about how to support users if technology solutions fail, and ethical issues about how to manage emotional connection and attachment to more anthropomorphic solutions. It illustrates that technology represents part of the solution but not the entire solution [70], and that service providers must balance the need for human connection with needs to increase efficiency and reduce demand on health services.

Our review found that participants experience a variety of usability and usage issues, and that flexibility of use and ability to configure a solution increases ease of use. Older people with MCI feel more comfortable, confident and in control when solutions are consistent, predictable and integrate into their existing lifestyles and routines. People often use multiple toolsets and technology providers should prioritize consistency and seamless integration of their products. Overall, our analysis reveals that technology solutions should support how people behave and live, not require them to change and adapt to how technology operates.

Some studies recommend that technology solutions for older people with MCI should be designed together with this population [43], [45], [87], [91], [100], [102], [128] to ensure that all needs are identified and appropriately prioritized. This recommendation is supported by our findings which shows that while there is consistency in the top priority need, there are differences in how participants in different locations prioritize the order of needs. Existing research shows that there is cultural and contextual influence on TA, and differences in the needs and priorities for independent living of older people in Sweden and Poland, and the technologies they would consider using [129]. Given the evidence for the importance of perceived usefulness and ease of use as factors for TA, it is critical to understand the needs and characteristics of the target user base so we can design to maximize TA. User Centered Design (UCD), first proposed by Kling [130], is a design methodology which focuses on usability and UX. UCD promotes inclusion of end users in every stage of solution creation, starting with initial data gathering to understand users' characteristics, environment of use and requirements. UCD is iterative and incorporates regular artefact evaluation and refinement based on user feedback. UCD processes and techniques have been applied successfully to design assistive technologies for people with dementia and shown to be an effective way to accurately capture requirements and ensure solution usability and acceptance [131].

## Scope of future work

This review revealed that older people with MCI have opinions about how and when they want to use technology, and it is imperative that future solutions are designed collaboratively with this population. Research is needed to accurately identify the needs of older people with MCI so we understand what would be considered useful, and how best to position technology in their lives. It would also be beneficial to identify which needs are common and which needs vary depending on demographics, location or other factors.

Solutions designed for older people with MCI must generate a positive experience because this affects a user's comfort, confidence and overall well-being. This review collated suggestions from participants, but more research is required to fully understand the factors to be considered.

Participant feedback suggests that personalization and control/enablement are important for older people with MCI, but it is not clear how choice with respect to technology, device and interaction modality affects TA. This review identified that older people with MCI do have preferred devices and interaction modalities, however further work is needed to fully understand effectiveness of different combinations of interaction modalities and how these are influenced by effects of MCI, context of use, device being used and task being performed.

Based on the findings, older people with MCI want to have autonomy, independence, and feel safe. They are accepting of technology if it positively contributes to their lifestyle and provides a sense of well-being. Future research should aim to provide clarity about how to effectively meet safety, privacy and security needs, and establish technology as an enabler for independence and autonomy in older age.

Finally, collection of longitudinal data about technology use by older people with MCI would capture changes to usage patterns, preferences and adoption factors as MCI advances, and enable us to understand how solution designs should allow for adaption to support MCI progression.

## Study strengths and limitations

This scoping review comprehensively searched nine databases for studies over a 10-year period. It considered research about a broad set of technology solutions proposed for older people with MCI, in the pre- and post-Covid-19 environments. However, the review is not without limitations. It only included studies in English, so it is possible untranslated relevant studies were overlooked. Most studies published their outcomes promptly after data collection, however [72] began data gathering in 2016 but did not publish the results until 2022. This is a long time in terms of technology evolution and the potential for change in peoples' opinions and attitudes. Some studies did not analyze non-verbal feedback, so it is possible that important information was missed [82], [85].

Most studies established an MCI diagnosis using the Montreal Cognitive Assessment (MoCA) [132] or the Mini-Mental State Examination (MMSE) [133] or Petersen's criteria [134]. However, the cut-off criteria for each assessment tool were not applied consistently across studies.

The definition of 'older adult' is not consistent across the 83 studies. Most studies considered 65+ as 'older', some studies nevertheless included participants aged between 48-59 [53], [58], [82], [94], [135], [136], [137], [138]. Some studies did not have an equal representation of genders.

Many studies listed small participant numbers or a lack of cultural diversity as limitations, and most studies included participants from only one location, meaning it may not be possible to generalize outcomes. Less than half the studies were completed in a real-life setting, so outcomes may not be fully representative of real-life results.

Studies did not collect a consistent set of participant data (such as IT literacy, existing devices and technology being used, education level, or profession) so it is not possible to comment on how these may have impacted participants' opinions, although age, sex, education level or experience are listed as moderators in existing technology acceptance models such as TAM, UTAUT and MATOA.

Some studies did not have people with MCI represented in all phases or excluded people with physical limitations, so the opinions of these people have not been considered.

## Conclusions

This scoping review identified that improving or maintaining their current level of cognition is the priority need for older people with MCI. Proposed technology solutions are perceived as useful by this population, even though gaps exist where functional support is missing. However, they are not perceived as completely easy to use, due to a variety of usability issues. In addition, technology solutions may have a positive or negative effect on the user's comfort, confidence, self-esteem, self-image, and overall well-being, depending on the feelings and emotions triggered by the solution design. Older people with MCI have preferences for how they use and interact with technology, and which devices they use. They prefer multimodal interaction, especially speech, visual/text and touch, because it is effective and provides options for use, and devices which are comfortable, convenient, light, readily available, cost-effective, and have large screens. Future work should focus on (i) gathering information about needs of older people with MCI and clarifying how these are dependent on location, demographics or other factors (ii) improving ease of use and UX, (iii) enabling customization and personalization, (iv) further exploring interaction preferences and effectiveness of different interaction modes, (v) providing options for multimodal interaction, and (vi) integrating technology solutions more seamlessly into people's existing lifestyle and routines.

# Acknowledgements

The authors would like to thank University of Canberra librarian Murray Turner for his advice regarding databases to search.

# Funding

The first author wishes to gratefully acknowledge the support of her research by an 'Australian Government Research Training Program Scholarship'.

# Conflict of interest

The authors declare no conflict of interest.

# Abbreviations

| | |
|---|---|
| ECA | Embodied Conversational Agent |
| HCI | Human Computer Interaction |
| H-TAM | Healthcare Technology Acceptance Model |
| MATOA | Model for the Adoption of Technology by Older Adults |
| MCI | Mild Cognitive Impairment |
| MMSE | Mini-Mental State Examination |
| MoCA | Montreal Cognitive Assessment |
| PRISMA | Preferred Reporting Items for Systematic Reviews and Meta-Analyses |
| STAM | Senior Technology Acceptance Model |
| TA | Technology Adoption |
| TAM | Technology Acceptance Model |
| UTAUT | Unified Theory of Acceptance and Use of Technology |
| UCD | User Centered Design |
| UX | User Experience |
| VR | Virtual Reality |

# Appendix

**Table 2:** Summary of device and interaction modality for studies (*the greyed-out cells indicate studies where no device was used*)

| | Hand | | | | | | Foot | | Auditory | | Visual | | | Haptic | |
|---|---|---|---|---|---|---|---|---|---|---|---|---|---|---|---|
| **Authors** | **Touch screen** | **Keyboard or mouse** | **Controller** | **Joystick** | **Push** | **Gesture** | **Cycle** | **Pedal /rudder /mat** | **Speech (out only)** | **Speech (in/out)** | **Video** | **Facial expression on solution** | **Light display** | **Robot moves** | **Vibrate** |
| Afifi, T. et al. | | | | | | | | | | √ | | | | | |
| Bartels, S. L. et al. | √ | | | | | | | | | | | | | | |
| Beentjes, K. M. et al. | √ | | | | | | | | | | | | | | |
| Bernini, S. et al. | | √ | | | | | | | | | | | | | |
| Bernini, S. et al. | √ | | | | | | | | | | | | | | |
| Bogza, L. et al. | | √ | | | | | | | | | | | | | |
| Bouzida, A. et al. | √ | | | | | | | | √ | | | √ | | √ | |
| Chang, C. et al. | | | | | | √ | | | | | | | | | |
| Chen, K. et al. | √ | | | | | | | | | | | | | | |
| Christiansen, L. et al. | | | | | | | | | | | | | | | |
| Collette, B. et al. | √ | | | | | | | | | | | | | | |
| Contreras-Somoza, L. M. et al. | √ | | | | | | | | | | | | | | |
| Contreras-Somoza, L. M. et al. | √ | | | | | | | | | | | | | | |
| Cunnah, K. et al. | | √ | | | | | | | | | | | | | |
| Dekker-van Weering, M. G. H. et al. | √ | √ | | | | | | | √ | | | | | | |
| Demiris, G. et al. | √ | | | | | | | | | √ | √ | √ | | | |
| Dixon, E. et al. | | | | | | | | | | | | | | | |
| Franco-Martín, M. A. et al. | √ | √ | | | | | | | | | | | | | |
| Gasteiger, N. et al. | √ | | | | | | | | √ | | | | | | |
| Gelonch, O. et al. | √ | | | | | | | | | | | | | | |
| Guzman-Parra, J. et al. | | | | | | | | | | | | | | | |
| Haesner, M. et al. | | | | | | | | | | | | | | | |
| Hassandra, M. et al. | | | √ | | | | | | | | | | | | |
| Heatwole, S. and Kendra, S. | | | | | | | | | | | | | | | |
| Hedman, A. et al. | | | | | | | | | | | | | | | |
| Hedman, A. et al. | | | | | | | | | | | | | | | |
| Horn, B. et al. | √ | | | | | | | | | | | | | | |

| | Hand | | | | | | Foot | | Auditory | | Visual | | | Haptic | |
|---|---|---|---|---|---|---|---|---|---|---|---|---|---|---|---|
| **Authors** | Touch screen | Keyboard or mouse | Controller | Joystick | Push | Gesture | Cycle | Pedal /rudder /mat | Speech (out only) | Speech (in/out) | Video | Facial expression on solution | Light display | Robot moves | Vibrate |
| Hu, H. et al. | | | | | | | | | | | | | | | |
| Infarinato, F. et al. | √ | | | | | | | | | | | | | | |
| Irazoki, E. et al. | √ | | | | | | | | √ | | | | | | |
| Konstantinos, V. et al. | √ | | | | | | | | | | | | | | |
| Korchut, A. et al. | | | | | | | | | | | | | | | |
| Kubota, A. et al. | √ | | | | | | | | | √ | | | | | |
| LaMonica, H. M. et al. | | | | | | | | | | | | | | | |
| Law, M. et al. | √ | | | | | | | | √ | | | | | | |
| Lazarou, I. et al. | | | | | | | | | | | | | | | |
| Leese, M. et al. | √ | | | | | | | | | | | | | | |
| Lindqvist, E. et al. | | | | | | | | | | | | | | | |
| Madjaroff, G. and Mentis, H. | | | | | | | | | | | | | | | |
| Maier, A. M. et al. | √ | | | | | | | | | | | | | | √ |
| Manca, M. et al. | √ | | | | | | | | √ | | | √ | √ | √ | |
| Manera, V. et al. | | √ | | | | | | | | | | | | | |
| Mathur, N. et al. | √ | | | | | | | | | √ | | | | | |
| Matsangidou, M. et al. | | | | | | | | | √ | | | | | | |
| Mattos, M. K. et al. | | √ | | | | | | | | | | | | | |
| McCarron, H. R. et al. | √ | √ | | | | | | | | | | | | | √ |
| Mehrabian, S. et al. | √ | | | | | | | | | | | | | | |
| Mondellini, M. et al. | | | √ | | | | | | | | | | | | √ |
| Moro, C. et al. | √ | | | | | | | | √ | | | √ | √ | √ | |
| Mrakic-Sposta, S. et al. | | | √ | | | | √ | | √ | | | | | | |
| Nie, Q. et al. | √ | √ | | | | | | | | √ | √ | | | | |
| Nieto-Vieites, A. et al. | √ | | | | | | | | | | | | | | |
| Ortega Morán, J. et al. | √ | | | | | | | | | | | | | | |
| Park, C. et al. | √ | | | | | | | | √ | | √ | | | | |
| Park, J. et al. | | | √ | | | | | | | | | | | | |
| Piasek, J. et al. | √ | | | √ | | | | | √ | | | | | | |
| Pino, M. et al. | √ | | | | | | | | | √ | | | | | |
| Quintana, M. et al. | √ | | | | | | | | | | | | | | |
| Rossi, S. et al. | √ | | | | | | | | | | | | | √ | |
| Saini, J. et al. | | √ | | | | | | | | | | | | | |

| | Hand | | | | | | Foot | | Auditory | | Visual | | | Haptic | |
|---|---|---|---|---|---|---|---|---|---|---|---|---|---|---|---|
| **Authors** | Touch screen | Keyboard or mouse | Controller | Joystick | Push | Gesture | Cycle | Pedal /rudder /mat | Speech (out only) | Speech (in/out) | Video | Facial expression on solution | Light display | Robot moves | Vibrate |
| Scase, M. et al. | √ | | | | | | | | | | | | | | |
| Schaham, N. G. et al. | √ | | | | | | | | | | | | | | |
| Scheibe, M. et al. | √ | √ | √ | | | | | | | | | | | | |
| Shamir, D. et al. | √ | | | | | | | | | | | | | | |
| Shellington, E. M. et al. | √ | | | | | | | | | | | | | | |
| Shin, M. H. et al. | √ | √ | | | | | | | | | | | | | |
| Stogl, D. et al. | | | | | √ | | | | | | | | | | |
| Stramba-Badiale, C. et al. | | √ | | √ | | | | √ | √ | | | | | | |
| Tuena, C.et al. | | | √ | | | | | √ | √ | | | | | | |
| Van Assche, M. et al. | √ | | | | | | | | | √ | | | | | |
| Wargnier, P. et al. | √ | | | | | | | | | √ | | | | | |
| Wolf, D. et al. | | | | | | | | | | √ | | | | | |
| Wu, Y. et al. | | | | | | | | | | | | | | | |
| Wu, Y. et al. | √ | | | | | | | | | √ | √ | | | | |
| Yamazaki, R. et al. | | | | | | | | | | √ | | | √ | √ | |
| Yun, S. J. et al. | | | √ | | | | | | | | | | | | |
| Yurkewich, A. et al. | √ | √ | | | | | | | | √ | | | | | √ |
| Zafeiridi, P.et al. | | √ | | | | | | | | | | | | | |
| Zedda, E. et al. | √ | | | | | √ | | | | √ | | | | √ | |
| Zhang, B. and Gao, Y. | | | | | | | | | | | | | | | |
| Zhang, Q. and Liu, Y. | √ | | | | | | | √ | | √ | | | | | |
| Zhu, D. et al. | √ | | | | | | | | | √ | | | | | |
| Zubatiy, T. et al. | √ | | | | | | | | | √ | | | | | |